\documentclass[12pt,a4paper]{article}

\usepackage[english]{babel}

\usepackage[a4paper,top=2cm,bottom=2.54cm,left=1.91cm,right=1.91cm]{geometry}
\usepackage[style=phys, backend=biber, sortcites=false]{biblatex}
\usepackage{amsmath}
\allowdisplaybreaks
\usepackage{amssymb}
\usepackage{amsthm}
\usepackage{graphicx}
\usepackage[colorlinks=true, allcolors=blue]{hyperref}
\usepackage[title]{appendix}
\usepackage{mathrsfs}
\usepackage{amsfonts}
\usepackage{caption}
\usepackage{authblk}
\usepackage{csquotes}
\usepackage[T1]{fontenc} 
\usepackage{lipsum}
\usepackage{float}

\usepackage{setspace}
\usepackage{titlesec}
\titleformat{\section} 
  {\normalfont\Large\bfseries}{\thesection.}{1em}{}
  
\theoremstyle{plain}

\theoremstyle{definition}

\title{Generalized Algebra Grounded on Nonadditive Entropies}

\author[1,2]{Leandro Lyra Braga Dognini\thanks{Corresponding author. E-mail: \href{mailto:leandro.dognini@uerj.br}{leandro.dognini@uerj.br}.}}
\author[3,4,5]{Constantino Tsallis}

\affil[1]{\small Rio de Janeiro State University, Rua São Francisco Xavier 524, 20550-900, Rio de Janeiro, Brazil}
\affil[2]{\small Legislative Advisory, Federal Senate of Brazil, Praça dos Três Poderes, 70165-900, Brasília, Brazil}
\affil[3]{\small Centro Brasileiro de Pesquisas Físicas and National Institute of Science and Technology for Complex Systems, Rua Dr. Xavier Sigaud 150, 22290-180, Rio de Janeiro, Brazil}
\affil[4]{Santa Fe Institute, 1399 Hyde Park Road, 87501, Santa Fe, USA}
\affil[5]{Complexity Science Hub, Metternichgasse 8, 1030 Vienna, Austria}
\date{December 18, 2025}

\begin{document}
\maketitle
\begin{abstract}
\noindent The class of $N$-body complex systems with total number of microscopic states given by $W(N) \sim \nu^{N^\gamma}\;(\nu >1, \,\gamma > 0)$ can be thermostatistically handled with the nonadditive entropic functional $S_\delta(\{p_{i}\}) = k\sum_{i=1}^W p_i \Bigl(\ln \frac{1}{p_i} \Bigr)^\delta \;(\delta>0,\,S_1=S_{BG})$, $S_{BG}=k\sum_{i=1}^W p_i \ln \frac{1}{p_i}$ being the Boltzmann-Gibbs functional. Indeed, $S_{\delta=1/\gamma}(\{1/W(N)\})=k[\ln W(N)]^{\frac{1}{\gamma}} \propto N$, as mandated by thermodynamics. Another wide class is that with $W(N) \sim N^\rho\;(\rho>0)$ and a generalized statistical mechanics grounded on the nonadditive entropic functional $S_q(\{p_{i}\})=k\sum_{i=1}^W p_i \ln_q \frac{1}{p_i} \;(q\in \mathbb{R},\;S_1=S_{BG})$, with $\ln_q z =\frac{z^{1-q}-1}{1-q}\; (z\geq0,\;q\in\mathbb{R},\;\ln_1 z=\ln z)$, satisfactorily handles such systems with $q=1-1/\rho$. Furthermore, for this class, the size of the corresponding admissible phase space is characterized by $\ln_q (x\otimes_q  y) =\ln_q x + \ln_q y,\, x,y\geq1,\,q\leq 1$, and the $q$-product $x\otimes_q  y=[x^{1-q}+y^{1-q}-1]^{\frac{1}{1-q}}_{+}\;(x\otimes_1  y=xy)$ also leads to the definition of a $q$-algebra. The entropic functional $S_{q,\delta}(\{p_{i}\})=k\sum_{i=1}^W p_i \Bigl(\ln_q \frac{1}{p_i} \Bigr)^\delta\;(q\in\mathbb{R},\delta>0)$ unifies both cases above: $S_{q,1}=S_q$, $S_{1,\delta}=S_\delta$ and $S_{1,1}=S_{BG}$. In this paper, we generalize the $q$-algebra associated with $S_{q}$ to a new one associated with $S_{q,\delta}$, namely the $(q,\delta)$-algebra.
\end{abstract}

\section{Introduction}\label{sec1}
Boltzmann-Gibbs (BG) statistical mechanics
\cite{Boltzmann1872,Boltzmann1877,Gibbs1902} constitutes one of the pillars of contemporary theoretical physics. It is applicable under the assumptions of strong chaos (i.e., positive maximal Lyapunov exponent, for classical nonlinear dynamical systems), mixing and ergodicity, and it is grounded on the additive entropic functional whose simplest form, for discrete admissible microscopic configurations (i.e., discrete \textit{phase space}), is given by
\begin{equation}
S_{BG}(\{p_i\})=k\sum_{i=1}^W p_i \ln \frac{1}{p_i}, 
\label{BGentropy}
\end{equation}
with $\sum_{i=1}^W p_i=1$, where $k$ is a conventional positive constant chosen once for ever (typically $k=k_B$, $k_B$ being the Boltzmann constant, in physics, and $k=1$ in computational sciences), and $W\ge 1$. Its maximal value is attained for equal probabilities (i.e., $p_i=1/W,\, 1\le i\le W$) and is given by the celebrated expression
\begin{equation*}
S_{BG}=k\ln W \,, 
\end{equation*}
which connects the microscopic and macroscopic descriptions of physical systems. The functional (\ref{BGentropy}) is {\it additive} \cite{Penrose1970}, i.e., for all probabilistically independent systems $A$ and $B$, we have that
\begin{equation}
S_{BG}(\{p_i^A p_j^B\})=S_{BG}(\{p_i^A\})+S_{BG}(\{p_j^B\}),
\label{additive}
\end{equation}
for $i=1,2,\dots,W_A$ and $j=1,2,\dots,W_B$. The BG theory has, for 150 years, been satisfactorily validated in uncountable physical systems. However, it fails -- sometimes severely -- in wide classes of natural, artificial and social complex systems, thus opening the door for a generalized theory. Such generalized statistical mechanics (currently referred to as {\it nonextensive statistical mechanics}, or, simply, {\it $q$-statistics}) was in fact proposed in 1988 \cite{Tsallis1987}, grounded on the following {\it nonadditive} entropic functional
\begin{equation}
S_{q}(\{p_i\})=k\frac{1-\sum_{i=1}^Wp_i^q}{q-1} = k\sum_{i=1}^W p_i\ln_q \frac{1}{p_i} = -k\sum_{i=1}^W p_i^q \ln_q p_i =-k\sum_{i=1}^W p_i \ln_{2-q} p_i,
\label{qentropy}
\end{equation}
for $q\in\mathbb{R}$ and $\sum_{i=1}^W p_i=1$. Notice that $W\ge 1$ includes only events whose probability is nonvanishing (clearly, this restriction is only relevant for $q<0$). 

The {\it $q$-logarithm}, $\ln_{q}:[0,+\infty]\rightarrow [-\infty,+\infty]$, is defined as 
\begin{equation}
\ln_q z = \begin{cases}
    \frac{z^{1-q}-1}{1-q},\,\textrm{if }q\neq 1,\\
    \ln z,\,\textrm{if }q=1,
\end{cases} 
\end{equation}
and its inverse, the {\it $q$-exponential}, $\exp_{q}:\textbf{Img}\ln_{q}\rightarrow[0,+\infty]$, is given by
\begin{equation}
\exp_{q}z= \begin{cases}
    (1+(1-q)z)^{\frac{1}{1-q}} ,\,\textrm{if }q\neq 1,\\
    \exp z,\,\textrm{if }q=1,
\end{cases}  
\end{equation}
where $\textbf{Img}\ln_{q}\subseteq[-\infty,+\infty]$ denotes\footnote{Let us remind that, for a function $f:X\rightarrow Y$, $\textbf{Img}\, f=\{y\in Y \mid \exists\, x\in X, f(x)=y\}$.}
\begin{eqnarray}
    \textbf{Img}\ln_{q}=\begin{cases}
        \Bigl[\frac{-1}{1-q}, +\infty\Bigr],\, q<1,\\
        [-\infty,+\infty],\, q=1,\\
        \Bigl[-\infty,\frac{1}{q-1}\Bigr],\, q>1.
    \end{cases}
\end{eqnarray}

The functional (\ref{qentropy}) attains its maximum for equal probabilities, given by
\begin{equation*}
S_q\biggr(\biggr\{\frac{1}{W}\biggr\}\biggr)=k\ln_q W \,.    
\end{equation*}
Also, we verify that
\begin{equation}
S_{q}(\{p_i^A p_j^B\})=S_{q}(\{p_i^A\})+ S_{q}(\{p_j^B\})+\frac{1-q}{k} S_{q}(\{p_i^A\})S_{q}(\{p_j^B\}),
\label{nonadditive}
\end{equation}
for $i=1,2,\dots,W_A$ and $j=1,2,\dots,W_B$, and $S_{q}$ recovers $S_{BG}$ in the limit $(1-q)/k \to 0$. Furthermore, the functional $S_q$ is the unique to simultaneously be both trace-form and composable \cite{EncisoTempesta2017,Tsallis2023}.

Focusing on black holes and related cosmological phenomena, a different nonadditive entropic functional has also been introduced, namely \cite{Tsallis2009}
\begin{equation}
S_\delta(\{ p_i\})=k \sum_{i=1}^W p_i \Bigl(\ln \frac{1}{p_i}\Bigr)^\delta,    
\end{equation}
for $\delta>0$, whose maximal value is given by
\begin{equation}
S_\delta\biggr(\biggr\{\frac{1}{W}\biggr\}\biggr)=k( \ln W)^\delta \,.  
\label{maximaldeltaentropy}
\end{equation}

The functional $S_\delta$ is trace-form but not composable\footnote{Still, we highlight that another functional, also focusing on black holes and related systems, has been introduced in the literature \cite{Tempesta2020,TsallisJensen2025}
which, in contrast with $S_\delta$, is composable but not trace-form. In addition, its maximum value is the same as that indicated in (\ref{maximaldeltaentropy}).}, and recovers $S_{BG}$ in the limit $\delta \to 1$. The functionals $S_q$ and $S_\delta$ have been unified through the following one \cite{TsallisCirto2013}:
\begin{equation}
S_{q,\delta}(\{ p_i\})=k \sum_{i=1}^W p_i \Bigl(\ln_q \frac{1}{p_i}\Bigr)^\delta,    
\end{equation}
for $q\in\mathbb{R}$ and $\delta>0$, which yields $S_{q,1}=S_q$, $S_{1,\delta}=S_\delta$, and $S_{1,1}=S_{BG}$. 

The applicability of nonadditive entropies and their consequences has been profusely validated in the literature\footnote{See bibliography in \href{https://tsallis.cbpf.br/biblio.htm}{https://tsallis.cbpf.br/biblio.htm}.}, very specifically in cold atoms in dissipative optical lattices \cite{Renzoni, LutzRenzoni2013}, granular matter \cite{GheorghiuOmmenCoppens2003,Combe}, nonlinear dynamical systems at the edge of chaos \cite{TirnakliBorges2016,TirnakliTsallis2020b,BountisVeermanVivaldi2020,RuizTirnakliBorgesTsallis2017,beckgauss},  high energy collisions of elementary particles at CERN \cite{WaltonRafelski2000,WongWilk2013,WongWilkCirtoTsallis2015,deppmanfractal4,deppmanfokker2,baptistaairton}, cosmology \cite{TsallisCirto2013,JizbaLambiase2022,JizbaLambiase2023,SalehiPouraliAbedini2023,DenkiewiczSalzanoDabrowski2023,TsallisJensen2025}, long-range interactions in many-body Hamiltonian systems \cite{CirtoAssisTsallis2014,CirtoRodriguezNobreTsallis2018}, overdamped systems such as type-II superconductivity \cite{AndradeSilvaMoreiraNobreCurado2010,CasasNobreCurado2019}, turbulence \cite{Boghosian1996,DanielsBeckBodenschatz2004}, economics \cite{Borland2002a,Borland2002b, kaizoji2003,RuizMarcos2018}, earthquakes \cite{AntonopoulosMichasVallianatosBountis2014}, asymptotically scale-invariant networks \cite{nunesrole,samuraiproperties,samurailuciano,cinardisubmitted,rute,rute2}, slow chemical reactions through quantum tunneling \cite{wildnature,wildnature3}, neurosciences \cite{abramovEEG,abramovEEG2}, among many other complex systems.

To close the present introduction, let us mention that an interesting binary operation has long been introduced in the literature of $q$-statistics, namely the following {\it $q$-product} \cite{Nivanen2003,Borges2004}:
\begin{equation}
x \otimes_q y= [x^{1-q}+y^{1-q}-1]_{+}^{\frac{1}{1-q}},
\end{equation}
where $[z]_+=\max\{z,0\}$, for $z\in[-\infty,+\infty]$, $x,y\geq0$, and $x \otimes_1 y=xy$. This $q$-generalized product is abelian and, for $x,y\ge 0$ satisfying conditions that are detailed later on, it is associative and fulfills an important property, namely
\begin{equation}
\ln_q (x \otimes_q y) = \ln_q x + \ln_q y \,,
\end{equation}
which is closely related to the size of the phase space of complex systems. Also, this $q$-generalized product has been instrumental to appropriately generalize the Fourier transform so as to eventually $q$-generalize the Central Limit Theorem \cite{Umarov2008,Umarov2010}.
Moreover, it grounds an entire generalized algebra. Indeed, it can be shown to be, under conditions specified later on, distributive with regard to an abelian $q$-generalized sum \cite{Borges2022}. 

The main goal of the present paper is to further generalize the $q$-algebra based now on $S_{q,\delta}$ instead of its particular case $S_q$. This will hopefully open the door to an even more general Central Limit Theorem, which might also be applicable to black holes and related astrophysical and cosmological matters.

\section{The generalized algebra}\label{sec2}
We define the {\it $(q, \delta)$-logarithm}, $\ln_{q,\delta}:[0,+\infty]\rightarrow [-\infty,+\infty]$, $q\in\mathbb{R}$, $\delta>0$, as
\begin{eqnarray}
\ln_{q,\delta}z \equiv \begin{cases}\Bigl\vert\frac{z^{1-q}-1}{1-q} \Bigr\vert^\delta_{\pm},\,\textrm{if }q\neq1,\\
\vert\ln z\vert^{\delta}_{\pm},\,\textrm{if }q=1,
\end{cases}
\end{eqnarray}
 with $\vert z\vert^{\delta}_{\pm}=\textrm{sign}(z)\vert z\vert^{\delta}$, $z\in[-\infty,+\infty]$\footnote{To define the $(q,\delta)$-logarithm, we let
\begin{eqnarray*}
   \textrm{sign}(z)&=&\begin{cases}
       -1,\,z<0,\\
       0,\,z=0,\\
       1,\,z>0,
   \end{cases}
\end{eqnarray*}
and convention that $z+(+\infty)=(+\infty)+z=+\infty$, for $z\in\mathbb{R}$, $(+\infty)+(-\infty)=(-\infty)+(+\infty)=-\infty$, $0(+\infty)=(+\infty)0=0$, $1/0=+\infty$ and $1/+\infty=0$.}. One may directly notice that
\begin{eqnarray}
    \textbf{Img}\ln_{q,\delta}=\begin{cases}
        \Bigl[\frac{-1}{(1-q)^{\delta}}, +\infty\Bigr],\, q<1,\\
        [-\infty,+\infty],\, q=1,\\
        \Bigl[-\infty,\frac{1}{(q-1)^{\delta}}\Bigr],\, q>1,
    \end{cases}
\end{eqnarray}
and that
\begin{eqnarray}\label{eqImgEquivalence}
    z\in \textbf{Img}\ln_{q,\delta} \iff 1+\vert 1-q\vert^{\delta}_{\pm}z\geq0,
\end{eqnarray}
for $z\in[-\infty,+\infty]$. The inverse of $\ln_{q,\delta}(\cdot)$ is the {\it $(q, \delta)$-exponential}, $\exp_{q,\delta}:\textbf{Img}\ln_{q,\delta}\rightarrow [0,+\infty]$, given by
\begin{eqnarray}
    \exp_{q,\delta}z \equiv \begin{cases}(1+(1-q)\vert z\vert^{\frac{1}{\delta}}_{\pm})^{\frac{1}{1-q}},\,\textrm{if }q\neq1,\\
    \exp\vert z\vert^{\frac{1}{\delta}}_{\pm},\,\textrm{if }q=1,
    \end{cases}
\end{eqnarray}
so that $\exp_{q,\delta}(\ln_{q,\delta}z)=z$, for $z\geq0$, and $\ln_{q,\delta}(\exp_{q,\delta}z)=z$, for $z\in\mathbf{Img}\ln_{q,\delta}$. Also, for $x,y,(\vert x\vert^{\frac{1}{\delta}}_{\pm}+\vert y\vert^{\frac{1}{\delta}}_{\pm}+(1-q)\vert xy\vert^{\frac{1}{\delta}}_{\pm})^{\delta}\in\mathbf{Img}\ln_{q,\delta}$, we have
\begin{eqnarray*}
    (\exp_{q,\delta}x)\, (\exp_{q,\delta}y)
    &=&\exp_{q,\delta}((\vert x\vert^{\frac{1}{\delta}}_{\pm}+\vert y\vert^{\frac{1}{\delta}}_{\pm}+(1-q)\vert xy\vert^{\frac{1}{\delta}}_{\pm})^{\delta}).
\end{eqnarray*}

We verify {\it entropic nonadditivity}, more precisely
\begin{eqnarray*}
\ln_{q,\delta} (x y) 
=\biggr\vert \vert\ln_{q,\delta} x\vert^{\frac{1}{\delta}}_{\pm} + \vert\ln_{q,\delta} y\vert^{\frac{1}{\delta}}_{\pm} +(1-q)\vert\ln_{q,\delta} x \ln_{q,\delta} y\vert^{\frac{1}{\delta}}_{\pm}\biggr\vert^\delta_{\pm}\ne \ln_{q,\delta}x + \ln_{q,\delta} y,
\end{eqnarray*}
for $x,y\in(0,+\infty)$, $(q,\delta)\ne (1,1)$. However, the additivity of $S_{1,1}=S_{BG}$ derived from $\ln(xy)=\ln x +\ln y$, $x,y>0$, leads us to define, for $x,y\in[0,+\infty]$,
the {\it $(q,\delta)$-product}, $x \otimes_{q,\delta}  y$, as
\begin{eqnarray}
x \otimes_{q,\delta}  y &=& \begin{cases}
    \exp_{q,\delta}(\ln_{q,\delta} x + \ln_{q,\delta} y),\, \textrm{if } \ln_{q,\delta} x + \ln_{q,\delta} y \in \textbf{Img}\ln_{q,\delta},\\
    0,\, \textrm{if } \ln_{q,\delta} x + \ln_{q,\delta} y \notin \textbf{Img}\ln_{q,\delta},
\end{cases}
\end{eqnarray}
hence,
\begin{eqnarray*}
\ln_{q,\delta} (x \otimes_{q,\delta} y) = 
\ln_{q,\delta} x + \ln_{q,\delta} y ,
\end{eqnarray*}
when $\ln_{q,\delta} x + \ln_{q,\delta} y \in \textbf{Img}\ln_{q,\delta}$. Due to (\ref{eqImgEquivalence}),
\begin{eqnarray}\label{eqQDProduct}
\ln_{q,\delta} x + \ln_{q,\delta} y \in \textbf{Img}\ln_{q,\delta}
\iff 1+\vert 1-q\vert^{\delta}_{\pm}(\ln_{q,\delta}x+\ln_{q,\delta}y)\geq0,  
\end{eqnarray}
and, therefore, we can write
\begin{eqnarray}\label{generalizedProduct}
    x \otimes_{q,\delta}  y
    =\begin{cases}\Bigl(1+ \Bigl\vert\Bigl[1+\vert 1-q \vert^{\delta}_{\pm}(\ln_{q,\delta}x +\ln_{q,\delta}y)\Bigr]_{+}-1\Bigr\vert^{\frac{1}{\delta}}_{\pm}\Bigr)^{\frac{1}{1-q}},\,\textrm{if }q\neq1,\\
    \exp\Bigl\vert \vert\ln x\vert^{\delta}_{\pm}+\vert \ln y\vert^{\delta}_{\pm}\Bigr\vert^{\frac{1}{\delta}}_{\pm},\,\textrm{if } q=1,
    \end{cases}
\end{eqnarray}
with $[z]_{+}=\max\{z,0\}$, for $z\in[-\infty,+\infty]$. In particular,
\begin{eqnarray*}
   \ln_{q,\delta} (x \otimes_{q,\delta}  y) = \ln_{q,\delta} x + \ln_{q,\delta} y,
\end{eqnarray*}
for $x,y\geq0$ and $1+\vert 1-q\vert^{\delta}_{\pm}(\ln_{q,\delta}x+\ln_{q,\delta}y)\ge0$. The $(q,\delta)$-product behaves analogously in terms of the $(q,\delta)$-exponential, i.e., if $x,y, x+y\in\mathbf{Img}\ln_{q,\delta}$, then
\begin{eqnarray*}
    (\exp_{q,\delta}x) \otimes_{q,\delta} (\exp_{q,\delta} y) 
    &=& \exp_{q,\delta}(x+y).
\end{eqnarray*}

It is straightforward that the $(q,\delta)$-product is commutative and that
\begin{eqnarray}\label{eqQDProductNeutral}
    x \otimes_{q,\delta}  1 &=&x,
\end{eqnarray}
for $x\geq0$. Also,
\begin{eqnarray}\label{eqQDProductZero}
    x \otimes_{q,\delta} 0 &=&\begin{cases}
        \Bigl(1+ \Bigl\vert(x^{1-q}-1)^{\delta}-1\Bigr\vert^{\frac{1}{\delta}}_{\pm}\Bigr)^{\frac{1}{1-q}},\,\textrm{if }q<1\textrm{ and }x>1,\\
        0,\, \textrm{otherwise.}
    \end{cases}
\end{eqnarray}

For $x,y,z\ge0$, if
\begin{eqnarray}\label{conditionsProductAssoc}
    \{\ln_{q,\delta}x \otimes_{q,\delta} y + \ln_{q,\delta}z,\ln_{q,\delta}x + \ln_{q,\delta}y,\ln_{q,\delta}y + \ln_{q,\delta}z,\ln_{q,\delta}x + \ln_{q,\delta}y\otimes_{q,\delta} z\}\subset\mathbf{Img}\ln_{q,\delta},
\end{eqnarray}
then the $(q,\delta)$-product is associative. Indeed,
\begin{eqnarray}
    (x \otimes_{q,\delta} y) \otimes_{q,\delta}  z 
    = \exp_{q,\delta}(\ln_{q,\delta} x + \ln_{q,\delta} y + \ln_{q,\delta} z) = x \otimes_{q,\delta}  (y \otimes_{q,\delta}  z).
\end{eqnarray}
Consistently, we are allowed in this case to use the following simplified notation
\begin{equation*}
x \otimes_{q,\delta}  y \otimes_{q,\delta}  z 
\equiv (x \otimes_{q,\delta}  y) \otimes_{q,\delta}  z 
=x \otimes_{q,\delta}  (y \otimes_{q,\delta}  z) \,.
\end{equation*}

For $x,y\in[0,+\infty]$, the {\it $(q,\delta)$-sum}, $x \oplus_{q,\delta} y$, is given by
\begin{eqnarray}\label{generalizedSum}
    x \oplus_{q,\delta}  y &=& \begin{cases}
    \exp_{q,\delta}(h_{q,\delta}(x,y)),\, \textrm{if } h_{q,\delta}(x,y) \in \textbf{Img}\ln_{q,\delta},\\
    0,\, \textrm{if } h_{q,\delta}(x,y) \notin \textbf{Img}\ln_{q,\delta},
    \end{cases}
\end{eqnarray}
with the function $h_{q,\delta}:[0,+\infty]\times[0,+\infty]\rightarrow[-\infty,+\infty]$ still to be defined. One may directly see that the adequate definition for $q=\delta=1$ is $h_{1,1}(x,y)=\ln (x+y)$ and \textcite{Borges2022} have shown that, for $q\neq 1$,
\begin{eqnarray*}
    h_{q,1}(x,y)&=&\ln (\exp\ln_{q} x+\exp\ln_{q} y).
\end{eqnarray*}

Inspired by this fact, we heuristically define
\begin{eqnarray}
    h_{q,\delta}(x,y)=\ln(\exp \ln_{q,\delta}x + \exp \ln_{q,\delta}y).
\end{eqnarray}
Due to (\ref{eqImgEquivalence}),
\begin{eqnarray}\label{eqQDSumEquivalence}
    h_{q,\delta}(x,y)\in\mathbf{Img}\ln_{q,\delta} \iff 1+\vert 1-q\vert^{\delta}_{\pm}\ln(\exp \ln_{q,\delta}x + \exp \ln_{q,\delta}y)\ge0,
\end{eqnarray}
and, therefore, we can write
\begin{eqnarray}\label{eqQDSumExplicit}
    x \oplus_{q,\delta}  y &=&\begin{cases} \biggr(1+\Bigl\vert\Bigl[1+\vert 1-q\vert^{\delta}_{\pm}\ln(\exp \ln_{q,\delta}x + \exp \ln_{q,\delta}y)\Bigr]_{+}-1\Bigr\vert^{\frac{1}{\delta}}_{\pm}\biggr)^{\frac{1}{1-q}},\,\textrm{if }q\neq1,\\
    \exp\Bigl\vert\ln(\exp \vert \ln x\vert^{\delta}_{\pm}+\exp\vert\ln y\vert^{\delta}_{\pm})\Bigr\vert^{\frac{1}{\delta}}_{\pm},\,\textrm{if }q=1.
    \end{cases}
\end{eqnarray}
In particular,
\begin{eqnarray*}
    \exp\ln_{q,\delta}x\oplus_{q,\delta}y=\exp \ln_{q,\delta}x + \exp \ln_{q,\delta}y,
\end{eqnarray*}
for $x,y\geq0$ and $1+\vert 1-q\vert^{\delta}_{\pm}\ln(\exp \ln_{q,\delta}x + \exp \ln_{q,\delta}y)\ge0$.
It is straightforward that the $(q,\delta)$-sum is commutative and that
\begin{eqnarray}\label{eqQDSumNeutral}
    x\oplus_{q,\delta}0=\begin{cases} \biggr(1+\Bigl\vert\Bigl[1+\vert 1-q\vert^{\delta}_{\pm}\ln\Bigl(\exp \ln_{q,\delta}x + \exp \Bigl(\frac{-1}{(1-q)^{\delta}}\Bigr)\Bigr)\Bigr]_{+}-1\Bigr\vert^{\frac{1}{\delta}}_{\pm}\biggr)^{\frac{1}{1-q}},\,\textrm{if }q<1,\\
    x,\,\textrm{if }q\ge1,
    \end{cases}
\end{eqnarray}
for $x\geq0$. For $x,y,z\geq0$, if
\begin{eqnarray}\label{conditionsSumAssoc}
    \{h_{q,\delta}(x\oplus_{q,\delta}y,z),h_{q,\delta}(x,y),h_{q,\delta}(y,z),h_{q,\delta}(x,y\oplus_{q,\delta}z)\}\subset\mathbf{Img}\ln_{q,\delta},
\end{eqnarray}
then the $(q,\delta)$-sum is associative, since
\begin{eqnarray*}
    (x \oplus_{q,\delta}  y)\oplus_{q,\delta} z
    = \exp_{q,\delta}(\ln(\exp\ln_{q,\delta}x + \exp \ln_{q,\delta}y + \exp \ln_{q,\delta}z)) = x \oplus_{q,\delta}  (y\oplus_{q,\delta} z).
\end{eqnarray*}
Consistently, we are allowed in this case to use the following simplified notation
\begin{equation*}
x \oplus_{q,\delta}  y \oplus_{q,\delta}  z 
\equiv (x \oplus_{q,\delta}  y) \oplus_{q,\delta}  z 
=x \oplus_{q,\delta}  (y \oplus_{q,\delta}  z) \,.
\end{equation*}

Also, if
\begin{eqnarray}\label{conditionsSumDistributive}
    \{h_{q,\delta}(x\otimes_{q,\delta}y,x\otimes_{q,\delta}z),\ln_{q,\delta}x+\ln_{q,\delta}y,\ln_{q,\delta}x+\ln_{q,\delta}z,h_{q,\delta}(y,z),\dots\nonumber\\
    \dots \ln_{q,\delta}x+\ln_{q,\delta}(y\oplus_{q,\delta}z)\}\subset\mathbf{Img}\ln_{q,\delta},
\end{eqnarray}
then the $(q,\delta)$-sum is distributive, since
\begin{eqnarray*}
    (x \otimes_{q,\delta} y )\oplus_{q,\delta} (x\otimes_{q,\delta} z) 
    &=&\exp_{q,\delta}(\ln (\exp(\ln_{q,\delta} x + \ln_{q,\delta}y)+\exp(\ln_{q,\delta} x + \ln_{q,\delta} z))) \\
    &=& x \otimes_{q,\delta} (y\oplus_{q,\delta} z).
\end{eqnarray*}
In summary, $\mathcal{A}_{q,\delta}=\{[0,+\infty],\otimes_{q,\delta},\oplus_{q,\delta}\}$, $q\in\mathbb{R}$, $\delta>0$, is the algebra that generalizes, grounded on nonadditive entropies, the fundamental one $\mathcal{A}_{1,1}=\{[0,+\infty],\times,+\}$, with
\begin{eqnarray}
    x \otimes_{q,\delta}  y=\begin{cases}\Bigl(1+ \Bigl\vert\Bigl[1+\vert 1-q \vert^{\delta}_{\pm}(\ln_{q,\delta}x +\ln_{q,\delta}y)\Bigr]_{+}-1\Bigr\vert^{\frac{1}{\delta}}_{\pm}\Bigr)^{\frac{1}{1-q}},\,\textrm{if }q\neq1,\\
    \exp\Bigl\vert \vert\ln x\vert^{\delta}_{\pm}+\vert \ln y\vert^{\delta}_{\pm}\Bigr\vert^{\frac{1}{\delta}}_{\pm},\,\textrm{if } q=1,
    \end{cases}
\end{eqnarray}
and
\begin{eqnarray}
   x \oplus_{q,\delta}  y &=&\begin{cases} \biggr(1+\Bigl\vert\Bigl[1+\vert 1-q\vert^{\delta}_{\pm}\ln(\exp \ln_{q,\delta}x + \exp \ln_{q,\delta}y)\Bigr]_{+}-1\Bigr\vert^{\frac{1}{\delta}}_{\pm}\biggr)^{\frac{1}{1-q}},\,\textrm{if }q\neq1,\\
    \exp\Bigl\vert\ln(\exp \vert \ln x\vert^{\delta}_{\pm}+\exp\vert\ln y\vert^{\delta}_{\pm})\Bigr\vert^{\frac{1}{\delta}}_{\pm},\,\textrm{if }q=1,
    \end{cases}
\end{eqnarray}
for $x,y\in[0,+\infty]$. Figure \ref{fig1} illustrates these operations by plotting $(x\oplus_{q,\delta}x)/2$ and $\sqrt{x\otimes_{q,\delta}x}$, for $(q,\delta)=(1/2,3/2)$, and comparing them to $(x\oplus_{1,1}x)/2=x$ and $\sqrt{x\otimes_{1,1}x}=x$.

\begin{figure}[H]
\centering \includegraphics[width=\linewidth]{AlgebraGeneralized.jpeg}  \caption{Illustration of the $x$-dependence for $(q,\delta)=(1/2,3/2)$. {\it Left:}  $(x\oplus_{q,\delta}x)/2$ (solid red) and  $(x\oplus_{1,1}x)/2=x$ (dashed blue). {\it Right:} $\sqrt{x \otimes_{q,\delta}x}$ (solid red) and $\sqrt{x\otimes_{1,1}x}=x$ (dashed blue).}
\label{fig1}
\end{figure}

\section{Related algebraic structures}\label{sec2}
This section highlights algebraic structures related to the algebra $\mathcal{A}_{q,\delta}=\{[0,+\infty],\otimes_{q,\delta},\oplus_{q,\delta}\}$, $q\in\mathbb{R}$, $\delta>0$.

\subsection{The case $\mathcal{M}^{-}_{q,\delta}=\{[1,+\infty],\otimes_{q,\delta}\}$, for $q\leq 1$ and $\delta>0$}\label{subsecMonoid}
If $q\leq 1$ and $\delta>0$, then
\begin{eqnarray*}
    1+\vert x^{1-q}-1\vert^{\delta}_{\pm}+\vert y^{1-q}-1\vert^{\delta}_{\pm}=1+(x^{1-q}-1)^{\delta}+ (y^{1-q}-1)^{\delta}\geq0,
\end{eqnarray*}
for all $x,y\in[1,+\infty]$. Therefore, (\ref{eqQDProduct}) implies $\ln_{q,\delta} x + \ln_{q,\delta} y \in \textbf{Img}\ln_{q,\delta}$ and 
\begin{eqnarray*}
     x \otimes_{q,\delta}  y = \begin{cases}
         \Bigl(1+ \Bigl((x^{1-q}-1)^{\delta}+ (y^{1-q}-1)^{\delta}\Bigr)^{\frac{1}{\delta}}\Bigr)^{\frac{1}{1-q}},\,\textrm{if }q<1,\\
         \exp((\ln x)^{\delta}+(\ln y)^{\delta})^{\frac{1}{\delta}},\,\textrm{if } q=1,
         \end{cases}
\end{eqnarray*}
for all $x,y\in [1,+\infty]$. In particular, 
\begin{eqnarray}\label{idMonoid}
    \ln_{q,\delta} (x \otimes_{q,\delta} y) = \ln_{q,\delta} x + \ln_{q,\delta} y,
\end{eqnarray}
for all $x,y\in[1,+\infty]$. One may also notice that $x \otimes_{q,\delta} y\geq1$, for all $x,y\in[1,+\infty]$. Also, for $x,y,z\in[1,+\infty]$, conditions (\ref{conditionsProductAssoc}) are always satisfied and, therefore, $\otimes_{q,\delta}$ is associative. As a result of (\ref{eqQDProductNeutral}), $\mathcal{M}^{-}_{q,\delta}$ is an \textit{abelian monoid} \cite{Graetzer1979}. 

The reasoning that leads to this abelian monoid is closely related to the physical meaning of the $(q,\delta)$-product due to the following. It is a well-established definition that an entropic functional $S_{q,\delta}$ is \textit{additive} if, for \textit{independent} systems $A$ and $B$, with $W_{A},W_{B}\geq1$ the sizes of the corresponding phase spaces, one has
\begin{eqnarray}\label{eqAddit}
    S_{q,\delta}(\{p^{A}_{i}\})+S_{q,\delta}(\{p^{B}_{j}\})=S_{q,\delta}(\{p^{C}_{l}\}),
\end{eqnarray}
with $C=A+B$ the joint system. Since the definition itself states that the systems $A$ and $B$ are independent, this means that 
\begin{eqnarray}\label{eqTradPhaseSpace}
    W_{C}=W_{A}W_{B}=W_{A}\otimes_{1,1}W_{B},
\end{eqnarray}
and $p^{C}_{ij}=p^{A}_{i}p^{B}_{j}$, $1\leq i\leq W_{A}$, $1\leq j\leq W_{B}$. This independence assumption ensures that $S_{1,1}=S_{BG}$ satisfies (\ref{eqAddit}), although $S_{q,\delta}$, $(q,\delta)\ne(1,1)$, does not (leading to \textit{nonadditive} entropies), since 
\begin{eqnarray*}
S_{q,\delta}(\{p^{A}_{i}\})+S_{q,\delta}(\{p^{B}_{j}\})&=&k\sum^{W_{A}}_{i=1}\sum^{W_{B}}_{j=1}p^{A}_{i}p^{B}_{j} \, \ln_{q,\delta}\biggr(\frac{1}{p^{A}_{i}}\otimes_{q,\delta}\frac{1}{p^{B}_{j}}\biggr)\\
&\ne&k\sum^{W_{A}}_{i=1}\sum^{W_{B}}_{j=1}p^{A}_{i}p^{B}_{j} \, \ln_{q,\delta}\biggr(\frac{1}{p^{A}_{i}p^{B}_{j}}\biggr)\\
&=&S_{q,\delta}(\{p^{C}_{l}\}).
\end{eqnarray*}

However, we are precisely interested in the cases where the systems $A$ and $B$ are \textit{non-independent} in such a (strong) way that $S_{BG}$ is not thermodynamically extensive. In this setting of non-independent systems $A$ and $B$, if the phase spaces of $A$, $B$ and $C=A+B$ are equiprobable, then we can write
\begin{eqnarray}
    S_{q,\delta}\biggr(\biggr\{\frac{1}{W_{C}}\biggr\}\biggr)=S_{q,\delta}\biggr(\biggr\{\frac{1}{W_{A}}\biggr\}\biggr)+S_{q,\delta}\biggr(\biggr\{\frac{1}{W_{B}}\biggr\}\biggr)\iff W_{C}=W_{A}\otimes_{q,\delta}W_{B},
\end{eqnarray}
for $q\leq1,\delta>0$, with this equivalence well defined for all $W_{A},W_{B}\geq1$ due to (\ref{idMonoid}). Moreover, it is plausible that, if the equiprobability hypothesis is mildly relaxed, $W_{A}\otimes_{q,\delta}W_{B}$ still characterizes the size of the phase space of the complex joint system $C$.

Therefore, under an {\it equiprobability} assumption, the $(q,\delta)$-product is related to the \textit{size} of the phase space of the joint complex system $C=A+B$ when the respective entropies $S_{q,\delta}$ add up adequately, and the departure from the BG phase space size given by (\ref{eqTradPhaseSpace}) typically happens when long-range correlations become fundamental and the independence (or, nearly independence) hypothesis cannot be sustained.

Another interesting possibility related to the $(q,\delta)$-product cannot be excluded. Consider the case $C=A_1+A_2+ \dots + A_N=\sum_{n=1}^N A_n$, where the subsystems $\{A_n\}$ are all equal and not necessarily satisfy the equiprobability hypothesis. In such a case, it might happen that, in the $N\to\infty$ limit, the size of the phase space of $C$ is {\it asymptotically} characterized by $W_{A_1}\otimes_{q,\delta}W_{A_2} \otimes_{q,\delta} \dots  \otimes_{q,\delta}W_{A_N}$ for a special (possibly unique) pair $(q^\star, \delta^\star)$.

There is still an essential appointment to be made about $S_{q,\delta}$ and the concept of \textit{extensivity}. For systems $A_{1},\ldots,A_{N}$, $N\geq2$, we do {\it not} have in general
\begin{equation}
S_{q,\delta}\biggr(p^{\sum^{N}_{n=1}A_{n}}_{1},p^{\sum^{N}_{n=1}A_{n}}_{2},\dots,p^{\sum^{N}_{n=1}A_{n}}_{W_{\sum^{N}_{n=1}A_{n}}}\biggr)=\sum^{N}_{n=1}S_{q,\delta}(p^{A_{n}}_{1},p^{A_{n}}_{2},\ldots,p^{A_{n}}_{W_{A_{n}}}).
\end{equation}
Consequently, when the systems $A_{1},A_{2},\ldots, A_{N}$ are equal among them, we do {\it not} have in general
\begin{equation*}
S_{q,\delta}\biggr(p^{\sum^{N}_{n=1}A_{n}}_{1},p^{\sum^{N}_{n=1}A_{n}}_{2},\dots,p^{\sum^{N}_{n=1}A_{n}}_{W_{\sum^{N}_{n=1}A_{n}}}\biggr)=N\,S_{q,\delta}(p^{A_{1}}_{1},p^{A_{1}}_{2},\ldots,p^{A_{1}}_{W_{A_{1}}}).
\end{equation*}
However, it appears to be plausible that, for wide classes of systems (see, for instance, \textcite{CarusoTsallis2008}) and in the limit $N\to\infty$, a (typically unique) pair $(q^{*},\delta^{*})$ exists such that
\begin{eqnarray}
S_{q^{*},\delta^{*}}\biggr(p^{\sum^{N}_{n=1}A_{n}}_{1},p^{\sum^{N}_{n=1}A_{n}}_{2},\dots,p^{\sum^{N}_{n=1}A_{n}}_{W_{\sum^{N}_{n=1}A_{n}}}\biggr)\sim N\,S_{q^{*},\delta^{*}}(p^{A_{1}}_{1},p^{A_{1}}_{2},\ldots,p^{A_{1}}_{W_{A_{1}}})\propto N \,\,\,\,(N\to\infty),
\end{eqnarray}
as mandated by thermodynamics, and $S_{q^{*},\delta^{*}}$ is, in this case, \textit{extensive}. Finally, when the systems $A_{1},A_{2},\ldots, A_{N}$ are independent, notice that the additivity of $S_{1,1}=S_{BG}$ implies its extensivity.

\subsection{The case $\mathcal{S}_{q,\delta}=\{[1,+\infty],\oplus_{q,\delta}\}$, for $q\leq 1$ and $\delta>0$}\label{subsecSemigroup}
If $q\leq 1$ and $\delta>0$, then
\begin{eqnarray*}
    1+\vert 1-q\vert^{\delta}_{\pm}\ln(\exp \ln_{q,\delta}x + \exp \ln_{q,\delta}y)=1+(1-q)^{\delta}\ln(\exp \ln_{q,\delta}x + \exp \ln_{q,\delta}y)\ge0,
\end{eqnarray*}
for all $x,y\in[1,+\infty]$. Therefore, (\ref{eqQDSumEquivalence}) implies $h_{q,\delta}(x,y)\in \textbf{Img}\ln_{q,\delta}$ and
\begin{eqnarray*}
    x \oplus_{q,\delta}  y = \begin{cases}
        \biggr(1+(1-q)\Bigl(\ln\Bigl(\exp \Bigl(\Bigl(\frac{x^{1-q}-1}{1-q} \Bigr)^{\delta}\Bigr) + \exp \Bigl(\Bigl(\frac{y^{1-q}-1}{1-q} \Bigr)^{\delta}\Bigr)\Bigr)\Bigr)^{\frac{1}{\delta}}\biggr)^{\frac{1}{1-q}},\,\textrm{if }q<1,\\
    \exp\Bigl(\ln(\exp (\ln x)^{\delta}+\exp (\ln y)^{\delta})\Bigr)^{\frac{1}{\delta}},\,\textrm{if }q=1,
    \end{cases}
\end{eqnarray*}
for all $x,y\in[1,+\infty]$. One may notice that $x \oplus_{q,\delta} y\geq1$, for all $x,y\in[1,+\infty]$. Also, for $x,y,z\in[1,+\infty]$, conditions (\ref{conditionsSumAssoc}) are always satisfied and $\oplus_{q,\delta}$ is associative. Therefore, $\mathcal{S}_{q,\delta}=\{[1,+\infty],\oplus_{q,\delta}\}$ is an \textit{abelian semigroup}.

\subsection{The case $\mathcal{Z}_{q,\delta}=\{[1,+\infty],\otimes_{q,\delta},\oplus_{q,\delta}\}$, for $q\leq1$ and $\delta>0$}

By the same reasoning from Subsections \ref{subsecMonoid} and \ref{subsecSemigroup}, one may notice that, for all $x,y,z\in[1,+\infty]$, (\ref{conditionsSumDistributive}) is satisfied and distributivity holds. Also, $\mathcal{Z}_{q,\delta}=\{[1,+\infty],\otimes_{q,\delta},\oplus_{q,\delta}\}$ inherits the corresponding properties of the abelian monoid $\mathcal{M}_{q,\delta}$ and the abelian semigroup $\mathcal{S}_{q,\delta}$.

\subsection{The case $\mathcal{M}^{+}_{q,\delta}=\{[0,1],\otimes_{q,\delta}\}$, for $q\geq1$ and $\delta>0$}
If $q\geq 1$ and $\delta>0$, then
\begin{eqnarray*}
    1+\vert x^{1-q}-1\vert^{\delta}_{\pm}+\vert y^{1-q}-1\vert^{\delta}_{\pm}=1+(x^{1-q}-1)^{\delta}+ (y^{1-q}-1)^{\delta}\geq0,
\end{eqnarray*}
for all $x,y\in[0,1]$. Therefore, (\ref{eqQDProduct}) implies $\ln_{q,\delta} x + \ln_{q,\delta} y \in \textbf{Img}\ln_{q,\delta}$ and 
\begin{eqnarray*}
x \otimes_{q,\delta}  y = 
\begin{cases}
    \Bigl(1+ \Bigl((x^{1-q}-1)^{\delta}+ (y^{1-q}-1)^{\delta}\Bigr)^{\frac{1}{\delta}}\Bigr)^{\frac{1}{1-q}},\,\textrm{if }q>1,\\
    \exp-\Bigl((-\ln x)^{\delta}+(-\ln y)^{\delta}\Bigr)^{\frac{1}{\delta}},\,\textrm{if } q=1,    
\end{cases}
\end{eqnarray*}
for all $x,y\in [0,1]$. In particular, $\ln_{q,\delta} (x \otimes_{q,\delta} y) = 
\ln_{q,\delta} x + \ln_{q,\delta} y$. One may notice that $x \otimes_{q,\delta} y\leq1$, for all $x,y\in[0,1]$. Also, for $x,y,z\in[0,1]$, conditions (\ref{conditionsProductAssoc}) are always satisfied and, therefore, $\otimes_{q,\delta}$ is associative. Due to (\ref{eqQDProductNeutral}), $\mathcal{M}^{+}_{q,\delta}$, is an \textit{abelian monoid}.

\section{Conclusion}\label{sec3}
The $(q,\delta)$-product, given by (\ref{generalizedProduct}), has an identity given by (\ref{eqQDProductNeutral}), is commutative and, if conditions (\ref{conditionsProductAssoc}) are met, is associative. Moreover, if conditions (\ref{conditionsSumDistributive}) are satisfied, it is distributive with respect to the $(q,\delta)$-sum given by (\ref{eqQDSumExplicit}), which in turn is commutative and, if conditions (\ref{conditionsSumAssoc}) are satisfied, is associative. In addition, the $(q,\delta)$-sum has an identity for $q\geq1$ due to (\ref{eqQDSumNeutral}). 

In other words, in this paper we have introduced, on the grounds of the nonadditive entropic functional $S_{q,\delta}$, a {\it commutative algebra} $\mathcal{A}_{q,\delta}=\{[0,+\infty],\otimes_{q,\delta},\oplus_{q,\delta}\}$, $q\in\mathbb{R}$, $\delta>0$, that generalizes the fundamental one $\mathcal{A}_{1,1}=\{[0,+\infty],\times,+\}$ and whose $(q,\delta)$-product characterizes the size of the phase space of complex systems. 

This opens the door for various potentially powerful possibilities such as: (i) the introduction of a $(q,\delta)$-generalized Fourier transform; (ii) the proof of a $(q,\delta)$-generalized Central Limit Theorem (CLT); (iii) the definition of a $(q,\delta)$-generalized convolution product; (iv) the $(q,\delta)$-generalization of the Boltzmann-Gibbs statistical mechanics which would succeed in preserving the Legendre-transform structure of classical thermodynamics; and, last but not least, (v) we have perhaps now the seed for constructing the concept of a $(q,\delta)$-generalized vector-space\footnote{As an illustration of what we have in mind, focus on the {\it nonlinear} ordinary differential equation $\frac{dy}{dx^{\delta}}= ay^{q}$, $y(0)\geq0$, $a,q\in\mathbb{R}$, $x \ge 0$. Its general solution is given by 
\begin{eqnarray*}
y(x)&=& y(0)\otimes_{q} \exp_{q}(ax^{\delta})=y(0)\otimes_{q,1} \exp_{q,\delta}(a^{1/\delta}x).
\end{eqnarray*}

The case $y(0)=\delta=1$, $q \ge 1$, $a<0$, $x\ge 0,$ was applied to a biophysical system in 
\textcite{TsallisBemskiMendes1999}.}.

The proof of a $(q,\delta)$-generalized CLT would constitute a strong  mathematical basis for understanding the plethoric success of  the nonadditive entropic functionals $S_q$ \cite{Renzoni, LutzRenzoni2013,GheorghiuOmmenCoppens2003,Combe,TirnakliBorges2016,TirnakliTsallis2020b,BountisVeermanVivaldi2020,RuizTirnakliBorgesTsallis2017,beckgauss,WaltonRafelski2000,WongWilk2013,WongWilkCirtoTsallis2015,deppmanfractal4,deppmanfokker2,baptistaairton,CirtoAssisTsallis2014,CirtoRodriguezNobreTsallis2018,CirtoAssisTsallis2014,CirtoRodriguezNobreTsallis2018,AndradeSilvaMoreiraNobreCurado2010,CasasNobreCurado2019,Boghosian1996,DanielsBeckBodenschatz2004,Borland2002a,Borland2002b, kaizoji2003,RuizMarcos2018,AntonopoulosMichasVallianatosBountis2014,nunesrole,samuraiproperties,samurailuciano,cinardisubmitted,rute,rute2,wildnature,wildnature3,abramovEEG,abramovEEG2} and $S_\delta$ \cite{TsallisCirto2013,JizbaLambiase2022,JizbaLambiase2023,SalehiPouraliAbedini2023,DenkiewiczSalzanoDabrowski2023,TsallisJensen2025} in natural sciences. 

The $(q,\delta)$-generalization of the Boltzmann-Gibbs theory would constitute a formidable step forward towards understanding the deep connections between the microscopic, mesoscopic, and macroscopic worlds in nature, very specifically in the realm of natural, artificial, and social complex systems.  

Finally, $(q,\delta)$-generalized vector spaces could lead to the use of $(q,\delta)$-exponentials as bases in such spaces, which could in principle dramatically improve the performance of computational methods in sciences such as theoretical chemistry \cite{AndradeMundimMalbouisson2005,AndradeMalbouissonMundim2006}, global-optimization algorithms \cite{TsallisStariolo1996}, engineering \cite{GrecoTsallisRapisardaPluchinoContrafatto2020,MorshedyAlshammaiTyagiElbatalHamedEliwa2021,VinciguerraGrecoPluchinoRapisardaTsallis2023}, processing of images and signals \cite{PlastinoRosso2005,MohanalinBeenamolKalraKumar2010,DinizMurtaBrumAraujoSantos2010,ShiLiMiaoHu2012,TsallisTirnakli2020,AzawiSaidiJalabKahtanIbrahim2021,abramovEEG,abramovEEG2} and others.

\section*{Acknowledgments}
  Fruitful discussions with E. P. Borges, E. M. F. Curado,  H. S. Lima and C. K. Mundim are gratefully acknowledged. One of us (C. T.) also acknowledges partial financial support from CNPq and FAPERJ (Brazilian agencies).

\appendix
\section*{Appendix}\label{appx}
We state in this appendix the detailed algebraic derivations of the paper, specifically of (\ref{eqImgEquivalence}), (\ref{generalizedProduct}), (\ref{eqQDProductNeutral}), (\ref{eqQDProductZero}), (\ref{eqQDSumExplicit}) and (\ref{eqQDSumNeutral}). To obtain (\ref{eqImgEquivalence}), let $q<1$, $\delta>0$, so that
\begin{eqnarray*}
    1+\vert 1-q\vert^{\delta}_{\pm}z=1+(1-q)^{\delta}z\geq0 \iff z\geq \frac{-1}{(1-q)^{\delta}}\iff z\in \textbf{Img}\ln_{q,\delta},
\end{eqnarray*}
for $z\in[-\infty,+\infty]$. For $q>1$,
\begin{eqnarray*}
    1+\vert 1-q\vert^{\delta}_{\pm}z=1-(q-1)^{\delta}z\geq0 \iff z\leq \frac{1}{(q-1)^{\delta}}\iff z\in \textbf{Img}\ln_{q,\delta},
\end{eqnarray*}
for $z\in[-\infty,+\infty]$. Finally, for $q=1$,
\begin{eqnarray*}
    1+\vert 1-q\vert^{\delta}_{\pm}z=1\geq0
\end{eqnarray*}
for all $z\in[-\infty,+\infty]=\textbf{Img}\ln_{1,\delta}$, due to the convention that $0(+\infty)=0$.

To obtain (\ref{generalizedProduct}), notice that, if $q\neq1$, $1+\vert 1-q \vert^{\delta}_{\pm}(\ln_{q,\delta} x + \ln_{q,\delta} y)\ge 0$, then
\begin{eqnarray*}
    x \otimes_{q,\delta}y
    &=&(1+(1-q)\vert \ln_{q,\delta} x + \ln_{q,\delta} y\vert^{\frac{1}{\delta}}_{\pm})^{\frac{1}{1-q}}\\
    &=&\Bigl(1+\Bigl\vert \vert 1-q \vert^{\delta}_{\pm}(\ln_{q,\delta} x + \ln_{q,\delta} y)\Bigr\vert^{\frac{1}{\delta}}_{\pm}\Bigl)^{\frac{1}{1-q}}\\
    &=&\Bigl(1+\Bigl\vert \Bigr[1+\vert 1-q \vert^{\delta}_{\pm}(\ln_{q,\delta} x + \ln_{q,\delta} y)\Bigr]_{+}-1\Bigr\vert^{\frac{1}{\delta}}_{\pm}\Bigl)^{\frac{1}{1-q}}.
\end{eqnarray*}
Also, if $1+\vert 1-q \vert^{\delta}_{\pm}(\ln_{q,\delta} x + \ln_{q,\delta} y)<0$, then
\begin{eqnarray*}
    \Bigl(1+\Bigl\vert \Bigr[1+\vert 1-q \vert^{\delta}_{\pm}(\ln_{q,\delta} x + \ln_{q,\delta} y)\Bigr]_{+}-1\Bigr\vert^{\frac{1}{\delta}}_{\pm}\Bigl)^{\frac{1}{1-q}}=(1+\vert-1\vert^{\frac{1}{\delta}}_{\pm})^{\frac{1}{1-q}}=0=x \otimes_{q,\delta}y.
\end{eqnarray*}
We conclude that, for $q\neq1$,
\begin{eqnarray*}
    x \otimes_{q,\delta}y=\Bigl(1+\Bigl\vert \Bigr[1+\vert 1-q \vert^{\delta}_{\pm}(\ln_{q,\delta} x + \ln_{q,\delta} y)\Bigr]_{+}-1\Bigr\vert^{\frac{1}{\delta}}_{\pm}\Bigl)^{\frac{1}{1-q}}.
\end{eqnarray*}
If $q=1$, then
\begin{eqnarray*}
    x \otimes_{q,\delta}y=\exp_{q,\delta}(\ln_{q,\delta} x + \ln_{q,\delta} y)=\exp\Bigl\vert \vert\ln x\vert^{\delta}_{\pm}+\vert \ln y\vert^{\delta}_{\pm}\Bigr\vert^{\frac{1}{\delta}}_{\pm}.
    \end{eqnarray*} 

To obtain (\ref{eqQDProductNeutral}), notice that
\begin{eqnarray*}
    x \otimes_{q,\delta}  1 
    &=& \begin{cases}\Bigl(1+ \Bigl\vert\Bigl[1+\vert x^{1-q}-1\vert^{\delta}_{\pm}+\vert 1^{1-q}-1\vert^{\delta}_{\pm}\Bigr]_{+}-1\Bigr\vert^{\frac{1}{\delta}}_{\pm}\Bigr)^{\frac{1}{1-q}},\,\textrm{if }q\neq1,\\
    \exp\Bigl\vert \vert\ln x \vert^{\delta}_{\pm}+\vert\ln 1 \vert^{\delta}_{\pm}\Bigr\vert^{\frac{1}{\delta}}_{\pm},\,\textrm{if }q=1,
    \end{cases}\\
    &=&\begin{cases}
        \Bigl(1+ \Bigl\vert\vert x^{1-q}-1\vert^{\delta}_{\pm}\Bigr\vert^{\frac{1}{\delta}}_{\pm}\Bigr)^{\frac{1}{1-q}},\,\textrm{if }q\neq1,\\
        \exp\ln x,\,\textrm{if }q=1,
    \end{cases}\\
    &=&x,
\end{eqnarray*}
for $x\geq0$. 

To obtain (\ref{eqQDProductZero}), notice that
\begin{eqnarray*}
    x \otimes_{q,\delta}  0 
    &=&\begin{cases}
        \Bigl(1+ \Bigl\vert\Bigl[1+\vert x^{1-q}-1\vert^{\delta}_{\pm}+\vert-1\vert^{\delta}_{\pm}\Bigr]_{+}-1\Bigr\vert^{\frac{1}{\delta}}_{\pm}\Bigr)^{\frac{1}{1-q}},\,\textrm{if }q<1,\\
        \exp\Bigl\vert \vert\ln x \vert^{\delta}_{\pm}-\infty\Bigr\vert^{\frac{1}{\delta}}_{\pm} ,\,\textrm{if }q=1,\\
        \Bigl(1+ \Bigl\vert\Bigl[1+\vert x^{1-q}-1\vert^{\delta}_{\pm}+\vert +\infty-1\vert^{\delta}_{\pm}\Bigr]_{+}-1\Bigr\vert^{\frac{1}{\delta}}_{\pm}\Bigr)^{\frac{1}{1-q}},\,\textrm{if }q>1,
    \end{cases}\\
    &=&\begin{cases}
        \Bigl(1+ \Bigl\vert\Bigl[\vert x^{1-q}-1\vert^{\delta}_{\pm}\Bigr]_{+}-1\Bigr\vert^{\frac{1}{\delta}}_{\pm}\Bigr)^{\frac{1}{1-q}},\,\textrm{if }q<1,\\
        \exp\vert -\infty\vert^{\frac{1}{\delta}}_{\pm} ,\,\textrm{if }q=1,x<+\infty,\\
        \exp\vert +\infty-\infty\vert^{\frac{1}{\delta}}_{\pm} ,\,\textrm{if }q=1,x=+\infty,\\
        \Bigl(1+ \Bigl\vert\Bigl[1+\vert x^{1-q}-1\vert^{\delta}_{\pm}+\infty\Bigr]_{+}-1\Bigr\vert^{\frac{1}{\delta}}_{\pm}\Bigr)^{\frac{1}{1-q}},\,\textrm{if }q>1,
    \end{cases}\\
    &=&\begin{cases}
        \Bigl(1+ \vert-1\vert^{\frac{1}{\delta}}_{\pm}\Bigr)^{\frac{1}{1-q}},\,\textrm{if }q<1, x\le1,\\
        \Bigl(1+ \Bigl\vert(x^{1-q}-1)^{\delta}-1\Bigr\vert^{\frac{1}{\delta}}_{\pm}\Bigr)^{\frac{1}{1-q}},\,\textrm{if }q<1,x>1,\\
        0,\,\textrm{if }q=1,\\
        (+\infty)^{\frac{1}{1-q}},\,\textrm{if }q>1,
    \end{cases}\\
    &=&\begin{cases}
        \Bigl(1+ \Bigl\vert(x^{1-q}-1)^{\delta}-1\Bigr\vert^{\frac{1}{\delta}}_{\pm}\Bigr)^{\frac{1}{1-q}},\,\textrm{if }q<1\textrm{ and }x>1,\\
        0,\, \textrm{otherwise,}
    \end{cases}
\end{eqnarray*}
for $x\geq0$, due to the convention that $1/0=+\infty$, $1/(+\infty)=0$ and $+\infty+(-\infty)=-\infty$.

To obtain (\ref{eqQDSumExplicit}), notice that, if $q\neq 1$, $ 1+\vert 1-q\vert^{\delta}_{\pm}\ln(\exp \ln_{q,\delta}x + \exp \ln_{q,\delta}y)\ge0$, then
\begin{eqnarray*}
     x \oplus_{q,\delta} y&=&(1+(1-q)\vert\ln(\exp \ln_{q,\delta}x + \exp \ln_{q,\delta}y)\vert^{\frac{1}{\delta}}_{\pm})^{\frac{1}{1-q}}\\
     &=&\Bigl(1+\Bigl\vert\vert1-q\vert^{\delta}_{\pm}\ln(\exp \ln_{q,\delta}x + \exp \ln_{q,\delta}y)\Bigr\vert^{\frac{1}{\delta}}_{\pm}\Bigr)^{\frac{1}{1-q}}\\
     &=&\Bigl(1+\Bigl\vert\Bigl[1+\vert1-q\vert^{\delta}_{\pm}\ln(\exp \ln_{q,\delta}x + \exp \ln_{q,\delta}y)\Bigr]_{+}-1\Bigr\vert^{\frac{1}{\delta}}_{\pm}\Bigr)^{\frac{1}{1-q}}.
\end{eqnarray*}
Also, if $ 1+\vert 1-q\vert^{\delta}_{\pm}\ln(\exp \ln_{q,\delta}x + \exp \ln_{q,\delta}y)<0$, then
\begin{eqnarray*}
    \biggr(1+\Bigl\vert\Bigl[1+\vert 1-q\vert^{\delta}_{\pm}\ln(\exp \ln_{q,\delta}x + \exp \ln_{q,\delta}y)\Bigr]_{+}-1\Bigr\vert^{\frac{1}{\delta}}_{\pm}\biggr)^{\frac{1}{1-q}}=(1+\vert-1\vert^{\frac{1}{\delta}}_{\pm})=0=x \oplus_{q,\delta} y.
\end{eqnarray*}
We conclude that, for $q\neq1$,
\begin{eqnarray*}
    x \oplus_{q,\delta} y &=& \biggr(1+\Bigl\vert\Bigl[1+\vert 1-q\vert^{\delta}_{\pm}\ln(\exp \ln_{q,\delta}x + \exp \ln_{q,\delta}y)\Bigr]_{+}-1\Bigr\vert^{\frac{1}{\delta}}_{\pm}\biggr)^{\frac{1}{1-q}}.
\end{eqnarray*}
If $q=1$, then
\begin{eqnarray*}
    x\oplus_{q,\delta}y=\exp_{q,\delta}(\ln(\exp \ln_{q,\delta}x+\exp\ln_{q,\delta}y))=\exp\Bigl\vert\ln(\exp \vert \ln x\vert^{\delta}_{\pm}+\exp\vert\ln y\vert^{\delta}_{\pm})\Bigr\vert^{\frac{1}{\delta}}_{\pm}.
\end{eqnarray*}

Finally, to obtain (\ref{eqQDSumNeutral}), notice that
\begin{eqnarray*}
    x\oplus_{q,\delta}0
    &=&\begin{cases} \biggr(1+\Bigl\vert\Bigl[1+\vert 1-q\vert^{\delta}_{\pm}\ln\Bigl(\exp \Bigl(\Bigl\vert\frac{x^{1-q}-1}{1-q} \Bigr\vert^{\delta}_{\pm}\Bigr) + \exp \Bigl(\Bigl\vert\frac{-1}{1-q} \Bigr\vert^{\delta}_{\pm}\Bigr)\Bigr)\Bigr]_{+}-1\Bigr\vert^{\frac{1}{\delta}}_{\pm}\biggr)^{\frac{1}{1-q}},\,\textrm{if }q<1,\\
    \exp\Bigl\vert\ln(\exp \vert \ln x\vert^{\delta}_{\pm}+\exp(-\infty))\Bigr\vert^{\frac{1}{\delta}}_{\pm},\,\textrm{if }q=1,\\
    \biggr(1+\Bigl\vert\Bigl[1+\vert 1-q\vert^{\delta}_{\pm}\ln\Bigl(\exp \Bigl(\Bigl\vert\frac{x^{1-q}-1}{1-q} \Bigr\vert^{\delta}_{\pm}\Bigr) + \exp (-\infty)\Bigr)\Bigr]_{+}-1\Bigr\vert^{\frac{1}{\delta}}_{\pm}\biggr)^{\frac{1}{1-q}},\,\textrm{if }q>1,
    \end{cases}\\
    &=&\begin{cases} \biggr(1+\Bigl\vert\Bigl[1+\vert 1-q\vert^{\delta}_{\pm}\ln\Bigl(\exp \Bigl(\Bigl\vert\frac{x^{1-q}-1}{1-q} \Bigr\vert^{\delta}_{\pm}\Bigr) + \exp \Bigl(\frac{-1}{(1-q)^{\delta}}\Bigr)\Bigr)\Bigr]_{+}-1\Bigr\vert^{\frac{1}{\delta}}_{\pm}\biggr)^{\frac{1}{1-q}},\,\textrm{if }q<1,\\
    \exp\Bigl\vert\vert \ln x\vert^{\delta}_{\pm}\Bigr\vert^{\frac{1}{\delta}}_{\pm},\,\textrm{if }q=1,\\
    \biggr(1+\Bigl\vert\Bigl[1+\vert 1-q\vert^{\delta}_{\pm}\Bigl\vert\frac{x^{1-q}-1}{1-q} \Bigr\vert^{\delta}_{\pm}\Bigr]_{+}-1\Bigr\vert^{\frac{1}{\delta}}_{\pm}\biggr)^{\frac{1}{1-q}},\,\textrm{if }q>1,
    \end{cases}\\
    &=&\begin{cases} \biggr(1+\Bigl\vert\Bigl[1+\vert 1-q\vert^{\delta}_{\pm}\ln\Bigl(\exp \Bigl(\Bigl\vert\frac{x^{1-q}-1}{1-q} \Bigr\vert^{\delta}_{\pm}\Bigr) + \exp \Bigl(\frac{-1}{(1-q)^{\delta}}\Bigr)\Bigr)\Bigr]_{+}-1\Bigr\vert^{\frac{1}{\delta}}_{\pm}\biggr)^{\frac{1}{1-q}},\,\textrm{if }q<1,\\
    x,\,\textrm{if }q=1,\\
    \biggr(1+\Bigl\vert\vert x^{1-q}-1\vert^{\delta}_{\pm}\Bigr\vert^{\frac{1}{\delta}}_{\pm}\biggr)^{\frac{1}{1-q}},\,\textrm{if }q>1,
    \end{cases}\\
    &=&\begin{cases} \biggr(1+\Bigl\vert\Bigl[1+\vert 1-q\vert^{\delta}_{\pm}\ln\Bigl(\exp \Bigl(\Bigl\vert\frac{x^{1-q}-1}{1-q} \Bigr\vert^{\delta}_{\pm}\Bigr) + \exp \Bigl(\frac{-1}{(1-q)^{\delta}}\Bigr)\Bigr)\Bigr]_{+}-1\Bigr\vert^{\frac{1}{\delta}}_{\pm}\biggr)^{\frac{1}{1-q}},\,\textrm{if }q<1,\\
    x,\,\textrm{if }q\ge1,
    \end{cases}
\end{eqnarray*}
for $x\geq0$.


\printbibliography
\end{document}